\documentstyle[prb,twocolumn,aps]{revtex}

\input{epsf}

\begin{document}
\tolerance 10000

\draft

\title{A Critique of Two Metals}

\author{R. B. Laughlin}

\address{
Departrment of Physics\\
Stanford University\\
Stanford, California  94305 }

\twocolumn[
\date{ \today }
\maketitle
\widetext

\vspace*{-1.0truecm}

\begin{abstract}
\begin{center}
\parbox{14cm}{I argue that Anderson's identification of the conflict
between the fermi-liquid and non-fermi-liquid metallic states as the
central issue of cuprate superconductivity is fundamentally wrong.
All experimental evidence points to adiabatic continuability of the
strange metal into a conventional one, and thus to one metallic phase
rather than two, and all attempts to account theoretically for the
existence of a luttinger-liquid at zero temperature in spatial
dimension greater than 1 have failed.  I discuss the underlying reasons
for this failure and then argue that the true higher-dimensional
generalization of the luttinger-liquid behavior is a propensity of the
system to order.  This implies that the central issue is actually the
conflict between different kinds of order, i.e. exactly the idea
implicit in Zhang's paper.  I then speculate about how the conflict
between antiferromagnetism and superconductivity, the two principal
kinds of order in this problem, might result in both the observed
zero-temperature phase diagram of the cuprates and the luttinger-liquid
phenomenology, i.e. the breakup of the electron into spinons and holons
in certain regimes of doping and energy. The key idea is a quantum
critical point regulating a first-order transition between these
phases, and toward which one is first attracted under renormalization
before bifurcating between the two phases.  I speculate that this
critical point lies on the insulating line, and that the difference
between the Mott-insulator and fermi-liquid approaches to the
high-$T_c$ problem comes down to whether or not the superconducting
states made by n- and p-type doping can be continued into each other.
A candidate for the second fixed point required for distinct
superconducting phases is the P- and T-violating chiral spin liquid
state invented by me.}

\end{center}
\end{abstract}

\pacs{
\hspace{1.9cm}
PACS numbers: 71.10.Pm, 74.25.Dw, 74.20.Mn}
]

\narrowtext

In a recent paper Baskaran and Anderson \cite{bas} have criticized
Zhang's \cite{z} SO(5) theory of cuprate superconductivity on various
microscopic grounds following the general thinking of Greiter \cite{gr}
and also on the much more serious grounds that the entire idea of
ascribing the behavior of the cuprates to quantum criticality
\cite{qcp} is physically wrong.  The right idea, according to them,
is that a second kind of metallic state, the luttinger-liquid, is
present in the cuprates, and that the strange phenomenology of these
materials is due to the presence of this new state of matter
\cite{pwa}. The existence and importance of the non-fermi-liquid
state has been the central feature of Anderson's ideas on cuprate
superconductivity from the very beginning, and has had a powerful
influence on the development of the subject by virtue of being the only
genuinely new idea in the field.  But it is now obvious that we have
reached an impasse on this matter, and I think the controversy
surrounding Zhang's paper provides a much-needed opportunity to
question whether the conflict
between the fermi-liquid and the non-fermi-liquid might have been the
wrong issue. There are a great many reasons to be worried about
this.  What is the evidence that the non-fermi-liquid state is actually
different from the fermi-liquid in the sense of finite-temperature
adiabatic continuability?  Why is it so difficult to write down a
luttinger-liquid in spatial dimension greater than 1, much less find a
Hamiltonian that stabilizes such a state?  Why does existence of
the luttinger-liquid help identify the cause of cuprate
superconductivity?  What is the experiment that would resolve
the key controversies of the luttinger-liquid state in a definitive
way? There is still reason to take Anderson's phenomenological
observations seriously, in particular the interpretation of certain
experiments in terms of spinon and holon excitations into which the
electron decays, but there are also reasons to suspect that the
central issue he identified is not quite right. Zhang's ideas, which
are not completely right either in my view, have had the
salubrious effect of articulating an alternate view of the underlying
physics, namely the quantum criticality idea Baskaran and Anderson
are so quick to dismiss, in a particularly simple and elegant way using
equations that everyone can understand.  As a result there is now a
second important idea on the table, one that I think makes
considerably more sense than the luttinger-liquid idea, namely
that cuprate phenomenology might be fundamentally due to a conflict
between different kinds of {\it order}.

The antiferromagnetic and superconducting phases each derive,
according to Baskaran and Anderson, from a more fundamental
thermodynamic phase, the Mott insulator and the metal, respectively.
Let me for a moment defer the question of which metallic state is
intended here and concentrate on the existence of the Mott insulator,
a paramagnetic spin singlet with an energy gap for charged excitations
and no antiferromagnetic long-range order modeled after the ground
state of the Hubbard model at half-filling in 1 spatial dimension.
Baskaran and Anderson go further to say that the antiferromagnet is
a Mott insulator, and it is an antiferromagnet because it is a Mott
insulator, not vice versa; superexchange is a consequence of the
insulating state.  Unfortunately, ten years of work by some of the
best minds in theoretical physics have failed to produce {\it any}
formal demonstration of the existence of such a state at zero
temperature - essential here because everything conducts a little at
finite temperature - and dimension greater than 1.  Probably the
closest anyone came was my own work \cite{rbl} which produced a
state with a spin gap and discrete broken symmetries at the price of
long-range interactions, and which had a phenomenology inconsistent
with that of the cuprates. Anderson's views to the contrary, this
matters a great deal because one's inability to back up phenomenological
observations with a simple model that is easy to solve and makes sense
usually means that an important physical idea is either missing or
improperly understood. Another indicator that something is deeply
wrong is the inability of anyone to describe the elementary excitation
spectrum of the Mott insulator precisely even as pure phenomenology.
Nowhere can one find a quantitative band structure of the elementary
particle whose spectrum becomes gapped. Nowhere can one find precise
information about the particle whose gapless spectrum causes the
paramagnetism. Nowhere can one find information about the interactions
among these particles or of their potential bound state spectroscopies.
Nowhere can one find precise definitions of Mott insulator terminology.
The upper and lower Hubbard bands, for example, are vague analogues of
the valence and conduction bands of a semiconductor, except that they
coexist and mix with soft magnetic excitations no one knows how to
describe very well.

In light of the magnitude and scope of these problems it is rather
ironic that a zero-temperature state with {\it order} possessing all of
these properties, namely the conventional Hartree-Fock spin density
wave, has existed all along and can be written down and explained
easily.

Why is it so hard to construct a Mott insulating vacuum that makes
sense in 2 or more spatial dimensions when it can be done so readily
in 1?  I would like to address this question in the context of the pure
spin limit of the problem, as the difficulty is exhibited already
there, but the meaningfulness of this limit is not obvious and is one
of the things we need eventually to address. Consider a spin
Hamiltonian of the form

\begin{equation}
{\cal H} = \sum_{<j,k>} J_{jk} \; \vec{S}_j \cdot \vec{S}_k
\; \; \; ,
\end{equation}

\noindent
where $< \! j,k \! >$ denotes a sum over lattice pairs, not necessarily
near neighbors, and $J_{jk}$ is a translationally-invariant Heisenberg
exchange interaction of finite range. When the total spin per site is
integral it is possible to find exact solutions in any number of
dimensions that are legitimate spin liquids, in the sense of having
exponentially decaying correlations, an energy gap, and a common-sense
relationship between this gap and the correlation length \cite{s1}.
When the spin per site is half-integral, on the other hand, no such
solution has even been found, and such computer work as we have
indicates either order or inadequate sample-size convergence, i.e. that
the simulation is not large enough to determine one way or the other
whether ordering occurs.  This fundamental disparity between integral
and half-integral spins was anticipated by Lieb, Schultz, and Mattis
\cite{lieb} long before the discovery of high-$T_c$ superconductivity
and is manifested as the Haldane effect in 1 dimension \cite{fdm}. They
introduced the unitary operator

\begin{equation}
U = \exp \biggl\{ i \sum_j \; \frac{2\pi x_j}{L} S_j^z \biggr\}
\; \; \; ,
\end{equation}

\noindent
where $x_j$ denotes the x-coordinate of the $j^{th}$ lattice site and
$L$ denotes the sample size, which has the effect of rotating each spin
about the z-axis in a way that twists by $2\pi$ as one advances across
the sample.  This operator is defined in any number of dimensions, but
for the arguments to work properly in dimension greater than 1 it is
necessary to imagine a sample that is long and skinny, say 50
light-years wide and $10^5$ light-years long, and to have an odd number
of sites in the plane perpendicular to the long axis. Since $U$ rotates
all the spins in a given region together it is almost a symmetry
operator and therefore increases the expected energy by an amount that
vanishes as the sample size grows.  Denoting the exact ground state
by $| \Psi_0 \! >$, we have specifically

\begin{equation}
\frac{< \! \Psi_0 | U^\dagger {\cal H} U | \Psi_0 \! >}
     {< \! \Psi_0 | \Psi_0 \! >} -
\frac{< \! \Psi_0 | {\cal H} | \Psi_0 \! >}
     {< \! \Psi_0 | \Psi_0 \! >} \propto 1/L^2 \; \; \; ,
\end{equation}

\noindent
where $L$ denotes the sample length.  However, in a half-integral
spin system we also have

\begin{equation}
<\! \Psi_0 | U | \Psi_0 \! > = 0 \; \; \; ,
\end{equation}

\noindent
this following from the minus sign acquired by a spinor when it is
rotated by $2\pi$.  So $U | \Psi_0 \! >$ is exactly orthogonal to
$|\Psi_0 \! >$ when the spin per unit cell is half-integral. Since
$U$ does not conserve total spin, this implies that half-integral spin
systems have arbitrarily low-energy excitations in every spin
channel and are thus fundamentally infrared-degenerate.  This is
inconsistent with the energy gap characteristic of a legitimate quantum
spin liquid but an expected and necessary consequence of ordering.
So the simplest explanation of the computer experiments, the one I
believe to be right, is that half-integral spin systems have a
powerful propensity to order and do so almost always. The case of
1 dimension is an exception for the simple reason that continuous
symmetry breaking is impossible in 1 dimension.  The quantum spin
liquid in 1 dimension is {\it not} a new state of matter at all but
a stillborn antiferromagnet.  The higher-dimensional analogue of
the Haldane effect is the {\it non}existence of the Mott
insulator state at zero temperature.

It is very important to emphasize that this line of reasoning does
not contradict any of Anderson's phenomenological observations, but
simply contradicts the deeper physical meaning he assigns to them.
\cite{pwa} As a side effect they also relieve us of a great
intellectual burdon we should not have been carrying in the first place.
When one represents that a distinct quantum phase exists, one is not
allowed to adjust the Hamiltonian to make the desired behavior occur.
A good modeler starts from the correct equations, computes honestly,
and produces plots that match experiment. If, on the other
hand, one represents the behavior to be due to proximity to a
quantum critical point or line, it is {\it mandatory} to adjust the
Hamiltonian, as this is the only way to correctly identify the physical
principle unifying the behavior.  Furthermore it is quite possible for
the critical Hamiltonian to be ``unphysical'' in the sense of
containing parameters one would never find in nature.  The tasks of
demonstrating the existence of a phase and demonstrating the existence
of a critical surface have exactly opposite strategies and are mutually
incompatible.  So the pique shown by Baskaran and Anderson toward the
Zhang's work, particularly the lengths they go to criticize model
assumptions, specific values of parameters, computational strategies,
and so forth comes down to hostility toward the possibility
that the Mott insulator phase might not exist.

It is a tall order for anyone to demonstrate the existence of a phase
of matter without finding a transition to it.  I cannot, in fact, think
of a single instance in which this has been done.  If we were
considering a system at zero temperature the issue of two metallic
phases could be resolved very cleanly, as the Landau quasiparticle
either becomes arbitrarily well-defined as the energy scale is lowered
or it does not.  At zero temperature the luttinger-liquid could be
distinguished from the conventional fermi-liquid, and there would
have to be a phase phase transition between them.  However, in the
cuprates the temperature of the ostensible luttinger-liquid phase
- the normal state at optimal doping - cannot be lowered
to zero because superconductivity intervenes, so this test cannot be
applied.  This is unfortunate because above the superconducting dome
there is no evidence that this state cannot be continued
adiabatically into the metallic state at extreme overdoping. The
latter is thought by most of us to be conventional.  It is possible
that a critical point separating the two metallic phases exists
at zero temperature and is just covered up by the inopportune
occurrence of superconductivity, but I think this is incorrect.  It
requires the superconductivity to be unrelated to the more
important struggle between the two metallic states, and this is
inconsistent with the violent change in the carrier scattering rate
at the superconducting transition found by Bonn et al \cite{bonn} in
microwave skin conductivity experiments.

The continuability of the strange metal into the non-so-strange one
does, in fact, imply that starting from a weakly-interacting fermi sea
and summing Feynman graphs makes formal sense, and this implies that
fermi-liquid modeling also makes sense. It does not imply that this
is a {\it good} thing to do, however, because this approach relegates
all the strange phenomenology to the category of complicated detail,
whereas there is every reason to believe that some as-yet undiscovered
physical principle is at work. Anderson has, in fact, made an excellent
case for this.

The critical point idea could account for Anderson's phenomenology
quite completely if spinons and holons are the true elementary
excitations at the critical point. There are several reasons
for thinking this might be true, but all are necessarily indirect
because no critical point Hamiltonian with the requisite properties
has yet been discovered.

\begin{enumerate}

\item If spin-1/2 systems like to order then the critical point is the
      only place in the phase diagram where order does not occur.
      Absence of order is a sufficient condition, at least in a pure
      magnet, for spinons and holons to exist.

\item The two ordered phases in question have Goldstone modes which
      disperse linearly at long wavelengths.  This applies to all
      Hamiltonians in the basin of attraction of a given phase,
      including those arbitrarily close to the critical point. The
      modes in question do not, however, exist as sharp excitations
      at the critical point itself because there is no physical
      principle left preventing them from mixing. The critical
      Hamiltonian must therefore be characterized by a large number
      of strange low-lying excitations that can be organized by an
      arbitrarily small perturbation into these modes.  This occurs,
      for example, in the Hubbard model at half-filling, where ordinary
      electrons are organized by an arbitrarily small Hubbard U into
      the collective modes of either s-wave superconductivity or
      antiferromagnetism depending on the sign of U.  This is a
      somewhat unfortunate example because the conventional metal
      is usually understood to be phase, i.e. an attractive fixed
      point, which cannot be a critical point by definition.
      So we must have low-lying excitations that live at the critical
      point and nowhere else, that involve mixing of the Goldstones
      of the two phases, and are not conventional particles and holes.

\item There is reason to suspect that the principles of conformal
      symmetry can be abstracted to critical points in more than
      2 dimensions \cite{cardy}.  The assumptions of conformal
      invariance and dynamical scaling together lead to functional
      forms for response functions like those of the luttinger-liquid
      \cite{ma}.

\item Phenomenology suspiciously similar to that of the cuprates is
      observed in heavy-fermion materials under circumstances in
      which it can be attributed unambiguously to proximity to a
      critical point \cite{rosch}.

\end{enumerate}

\begin{figure}
\epsfbox{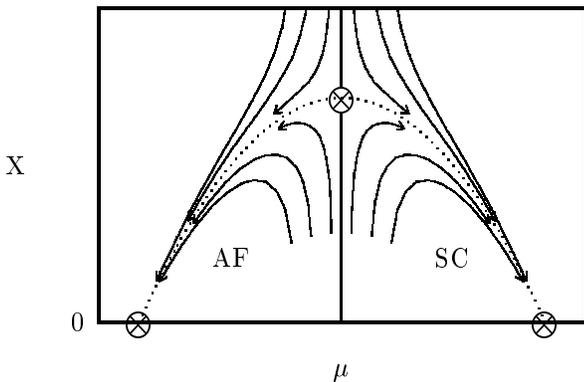}
\caption{Generic renormalization group flow for the proposed
         first-order line phase transition in the cuprates. X
         is an as-yet unidentified Hamiltonian parameter that is
         normally assumed to be zero. $\mu$ is the chemical potential.
         The bold line indicates the first-order surface.}
\end{figure}

Let me now proceed to speculate a bit on the nature of the
zero-temperature critical point or line which might be
responsible for the behavior Anderson has identified.
While there are many different kinds of order potentially present
in this system the {\it big} conflict is obviously between
superconductivity and antiferromagnetism.  I would therefore like to
assume that the critical point in question regulates the transition
between these, just as Zhang has done. I will also ignore
microscopics and start from the emperical fact that
both antiferromagnetic and d-wave superconducting order occur in
the cuprates and do, in fact, conflict. This is seen, for example,
in experiment in the advance with doping from Ne\`{e}l order to
``spin glass'' to superconductivity without intervention of a
normal-metal phase.  The spin glass is not a phase but a region of
increased sensitivity to disorder, i.e. exactly the kind of thing
one would expect to find at a first-order phase transition.  The
transition from isotropic antiferrmagnetism to superconductivity
must be first-order because the corresponding order parameters lie
in different irreducible representations of the lattice point group.
By elevating this conflict to a matter of importance we are, of
course, promoting the view that these two kinds of order have
the same microscopic cause.

In Fig. 1 I show a model renormalization group flow that might be
associated with such a transition.  I imagine that the first-order
transition extrapolates in Hamiltonian space into a surface, drawn
here as a line to emphasize the analogy with the quantum hall
transition \cite{rbl5}, terminating at a critical point.  The critical
Hamiltonian need not be experimentally accessible, and I have
accordingly labeled the second axis ``X''.  When the sample is small
there is no order and no interesting change to the behavior as doping
is varied.  As the sample size is increased, however, the low-energy
behavior flows toward that of the repulsive fixed point regulating
the transition, slows down there, and then finally bifurcates toward
the attractive fixed points characterizing the two phases.  The first
major effect seen as the sample size is increased is therefore not
the onset of order but the onset of criticality.  It is appropriate
to call this ``quantum disorder'' because the critical point
Hamiltonian is the only one that does not renormalize, i.e. does not
order.   The spectroscopic signatures of the critical point may also
be seen in the ordered phases by conducting experiments at {\it
intermediate momentum and energy scales}, as raising the energy scale
is equivalent to making the sample smaller and thus flowing backward
in Fig. 1.  It has been my view for quite some time that such reverse
scaling can be seen in many experiments, for example in the
photoemission ``spin gap'', the strange intermediate-energy
phenomenon characteristic of these materials which thus far has no
discoverable relevance to the superconductivity itself \cite{rbl2}.

\begin{figure}
\epsfbox{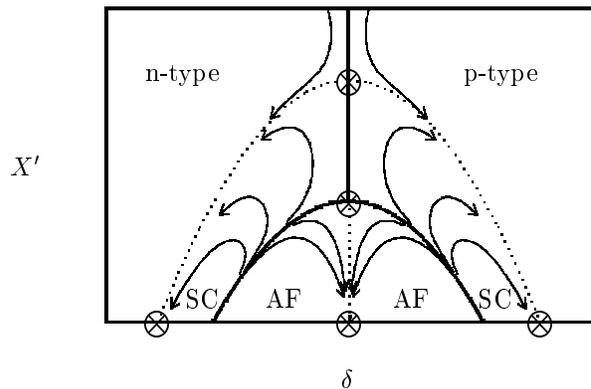}
\caption{Modified renormalization group flow for the cuprates
         that takes into account particle-hole symmetry, distinct
         superconducting states for n- and p-type doping, and a
         second fixed point representing a quantum-disordered
         state of the magnet.  Note that $\delta$ here means total
         oscillator strength below the ultraviolet cutoff, not the
         doping density.}
\end{figure}

Having identified a reasonable topology for the flow let us now
distort it, as shown in Fig. 2, so as to place the critical point
at zero doping. Baskaran and Anderson have correctly pointed out that
a continuous transition from superconductivity to antiferromagnetism
is not expected, even at a point, unless an additional physical
principle is at work. The physical effect I propose to exploit is the
vanishing of the superfluid density at half-filling. If the carrier
density becomes arbitrarily small then so do the superconducting order
parameter and the nonzero latent heat it usually necessitates. Locating
the critical point at the insulating boundary is also consistent with
the great body of RVB work, which always found an RVB vacuum, the
obvious prototype for the critical-point ground state, easy to define
at half-filling but nearly impossible to define for $\delta \neq 0$.
It is also relevant that the RVB states became disreputable precisely
because they showed signs of describing a quantum critical point
instead of a phase. Consider a wavefunction of the form

\begin{equation}
| \Psi \! > = P_\alpha | \Phi \! > \; \; \; ,
\label{gutz}
\end{equation}

\noindent
where $\Phi$ is a single-Slater-determinant electron wavefunction and

\begin{equation}
P_\alpha = \prod_j \biggl\{ 1 - \alpha \; n_{j \uparrow}
n_{j \downarrow} \biggr\} \; \; \; .
\label{disp}
\end{equation}

\noindent
So long as $0 \leq \alpha < 1$ the operator $P_\alpha$ defines an
invertible continuation of $| \Phi \! >$, albeit one that does not
preserve orthogonality. When $\alpha \rightarrow 1$, however, as
is the case in {\it all} the RVB wavefunctions, the map becomes
singular, and it becomes possible for states $| \Phi \! >$ with
fundamentally different symmetries to map to the same $| \Psi \! >$.
The most notorious example of this is the d-wave superconductor at
half-filling, which maps in this limit into a state with no
superconducting fluctuations at all - the vacuum most often referred
to as the RVB state.  This, however, is also the image of the s+id
superconducting state and the flux state, i.e. a Landau level on the
lattice \cite{kotliar,zou}. The behavior being described by these
wavefunctions is exactly that of a quantum critical point, i.e.
distinct phases of matter coming together at a single point in
parameter space and existing arbitrarily close to the point but not
at the point itself.

Fig. 2 differs from Fig. 1 in several important ways which are
relevant to the broader high-$T_c$ debate.  It is, of course,
reflection-symmetric about the zero-doping line, a property required
of any system with particle-hole symmetry, such as the Hubbard model
in the large-U limit.  The physical idea being expressed here is that
the system is like a semiconductor, and that the conducting states made
by n-type and p-type doping are mirror images of each other because the
quantum mechanics of the carriers is the same.  It is certainly the
case in the experimentally accessible parameter range that the cuprates
conduct only when doped and have the violently doping-dependent optical
sum rule expected of a semiconductor.  Fig. 2 also has {\it two}
superconducting states, one each for n-type and p-type doping, which
cannot be deformed into each other.  This is more controversial.
All experiments done to date on the cuprates have found two
superconducting states, but it is not clear that they are
adiabatically distinct.  Indeed many people believe that there is
only {\it one} superconducting state and that this is continuable
into a BCS state at half-filling.  Fig. 3 shows the flow expected
if this were the case.  I do not think this flow is right, but I
include it to make the dichotomy clear. Fig. 2 also differs from Fig. 1
in possessing a second fixed point above the first one to bifurcate
between the n-type and p-type regions.  This is required by
particle-hole symmetry if there are assumed to be two distinct
superconducting states, but more importantly by the physical idea
that the $\delta = 0$ line is insulating, for then the state to which
one flows at $\delta = 0$ cannot be a superconductor but must be
something else.  The absence of this second fixed point is the key
problem with Fig. 3.  Since everything above the dome in Fig. 3
conducts, we find that a miracle would be required to make the
superfluid density vanish at the critical point, which of
course was the reason for putting it at $\delta = 0$ in the first
place.  The superconductor-antiferromagnet transition would also
become synonymous with the metal-insulator transition, as occurs
in competition between spin density wave and BCS ground states in
a traditional metal, and we would be faced with the old problem of
explaining how such a transition could be continuous.   Finally,
our use of a spin model, such as that of Eq. (1), to motivate our
physical thinking would not be justified because it is valid only
within the basin of attraction of a phase containing that model.
A spin model can obviously never be attracted to a conducting fixed
point, so Fig. 3 would imply that no spin model could tell us
anything about the critical point, and also that the difference
between integral and half-integral spin is irrelevant because it is
invisible in the antiferromagnetically ordered state.

\begin{figure}
\epsfbox{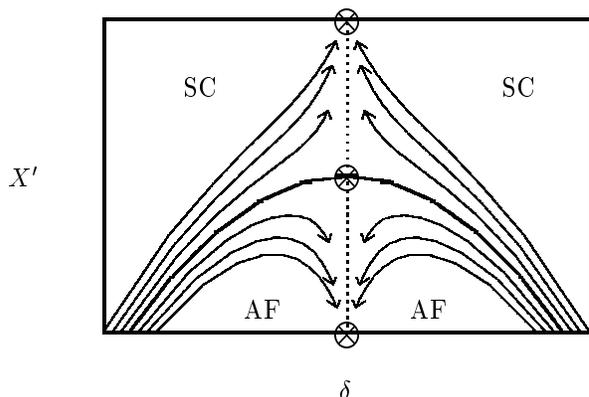}
\caption{Modified version of Fig. 2 appropriate to the case of only one
         superconducting state.  Note that this forces the
         superconductor-antiferromagnet transition to coincide with the
         metal-insulator transition and does not naturally lead
         to vanishing of the superfluid density.}

\end{figure}

Let us now consider the physical nature of the second fixed point.
At $\delta = 0$ it is attractive and represents a quantum phase of the
pure spin system different from the antiferromagnetically ordered one.
This phase must be characterized by some kind of order, as the relevant
spin model is half-integral, but the symmetry breaking can be only
discrete because the transition to it is second-order.  Spin-Peirels
order immediately comes to mind, particularly since it is known to
occur 1 dimension and to be associated with a scaling diagram similar
to the one I have drawn.  There is, however, no obvious reason for a
spin-Peirels state to be absolutely unstable to d-wave
superconductivity when doped, and this is essential for the system
to be a superconductor or an antiferromagnet, but nothing else, in the
thermodynamic limit.  For this reason I favor identifying the second
fixed point with the chiral spin liquid, the insulating state
characterized by short-range antiferromagnetic correlations, an
energy gap, and the P- and T-odd 3-spin order parameter $\vec{S}_i
\cdot (\vec{S}_j \times \vec{S}_k)$. The chiral spin liquid is
absolutely unstable to superconductivity when doped, by virtue of
the principles of anyon superconductivity \cite{rbl4}

It is with great reluctance that I introduce P and T violation to
this discussion. Cuprate superconductors do not appear to violate
T and P in the bulk, although they do so at surfaces \cite{greene},
and while the absence of such an effect in a real material can be
ascribed to the magnetic fields such a state would tend to generate,
the fact is that spontaneous breaking of P and T in the bulk has
become disreputable because it has not been seen experimentally.
Unfortunately the chiral spin liquid is the only insulating vacuum
known to be unstable to superconductivity for an identifiable
physical reason, and I believe that a reason is required, so I
continue to think that this vacuum is right despite the discouraging
experimental situation.  The implication is that there is a tendency,
perhaps only a subtle one, for the cuprates to break P and T
spontaneously, and that the conclusion that they do not is incorrect.
This would require yet another fixed point regulating the bifurcation
of the n-type and p-type superconducting states into right-handed
and left-handed versions.

The chiral spin liquid has always had the serious difficulty of
requiring long-range interactions to be stable.  This is a consequence
of the Lieb-Schultz-Mattis theorem, and it applies equally well to
the RVB vacuum, although it is less obvious in that case because an
antiferromagnetic vacuum with a small moment and a long correlation
length is a passable approximation to a state with no order and
power-law correlations.  However, this problem disappears once the
liquid becomes a repulsive fixed point rather than a phase, for then
it is no longer distinct from the superconducting state into which it
flows, and the requisite long-range forces can be attributed to
superfluid order.  The chiral spin liquid thus construed is
fundamentally different from the quantum-disordered state of
integral-spin systems as represented, say, by a nonlinear sigma model,
in that it cannot exist in isolation from its adjacent stabilizing
superfluid state. The superconductor into which the chiral spin liquid
flows has a small $d_{xy}$ order parameter superimposed on the usual
$d_{x^2 - y^2}$ one with a relative phase of $\pi / 2$, so that the
fermionic spectrum has an {\it energy gap}.  This gap measures the
amount of T-violation in the ground state and is the experimental
signature distinguishing such a state from a conventional d-wave
superconductor.  The relationship also works in reverse.  The action
of $P_\alpha$ for $\alpha = 1$ on this state at half-filling produces
the chiral spin liquid state.

I began this article with the proposition that the idea being defended
by Baskaran and Anderson \cite{bas} in their attack on Zhang \cite{z}
is fundamentally wrong because the luttinger-liquid does not exist
as a legitimate state of matter in spatial dimension greater than 1.
At the level of interpreting experiments this might be construed as
pedantry, for phenomenology based on a critical point is not so
different from phenomenology based on a phase if the energy resolution
is sufficiently crude.  However, this is not right because
high-$T_c$ is not the sort of problem in which modeling leads
inexorably to understanding.  We already know that the superconducting
state is deformable into a BCS state, albeit with d-wave symmetry, and
we also know that there is essentially no agreement with any of the
experimental minutiae usually cited as proof that the BCS theory is
correct. So the BCS paradigm is not the right one.  High-$T_c$ is
more like the strong interactions in that it presents us with an
abundance of experimental facts that are difficult to calculate from
first principles for known reasons, are for the most part unimportant,
and which require prioritization based on their ability to test
questions of principle.  In this context a misidentification of
the principle is the worst mistake one can possibly make, for it
causes unimportant things to be categorized as important and vice
versa.

Here is a summary of the important experimental implications of my
speculations:

\begin{enumerate}

\item There should be two distinct superconducting states, one each
      for n- and p-type doping, that cannot be deformed into each
      other without crossing a phase boundary.

\item There should be two and only two electronic phases -
      superconductivity and antiferromagnetism.  Striped
      phases count as antiferromagnetism.

\item The transition from superconductivity to antiferromagnetism as
      doping is reduced should be first-order, with no coexistence of
      the two kinds of order except through phase separation.  In
      other words, the two phases are antagonistic.

\item Luttinger-liquid behavior should be observable at lower and lower
      energy scales as the phase transition is approached from either
      direction and should persist to zero temperature at the
      transition.

\item There should be a tendency for the superconductor to develop a
      small $d_{xy}$ order parameter on top of the usual
      $d_{x^2 - y^2}$ one.

\end{enumerate}

I wish to thank S.-C. Zhang, M. Greiter, and E. Demler for numerous
helpful discussions and C. Henley and A. Auerbach for pointing out
two important errors in the original manuscript. This work was
supported primarily by the NSF under grant No. DMR-9421888.
Additional support was provided by the Center for Materials Research
at Stanford University and by NASA Collaborative Agreement NCC 2-794.

\section*{Afterword}

Shortly after this paper was released as a preprint Baskaran and
Anderson issued a reply to which I will now respond. \cite{bas2}.

Baskaran and Anderson begin by conceding my main point that the Mott
insulator ``is not a zero-temperature fixed-point'',  i.e. not a
legitimate state of matter at zero temperature, and then go on to
declare this unimportant because the zero-temperature constraint is too
``restricted''.  It is not too restricted.  If the Mott insulator
is not a phase at any temperature by any known definition then we must
either say {\it what} it is or stop writing papers about Mott
insulators.   Like Baskaran and Anderson, I believe that the
Mott phenomenon is a real effect, so I have attempted here to offer
ideas as to what it might be.  Declaring this to be a waste of time
is untenable in light of the situation in cuprate superconductivity,
and the matter is certainly not dealt with in Anderson's 1959 paper
\cite{pwa2}

The examples of Mott insulators given by Baskaran and Anderson -
CuCl$_2 \cdot$ 2H$_2$O, CuSO$_4 \cdot$ 5H$_2$O, iron oxides, and
hemoglobin - are all beautifully consistent with my views as they
are all adiabatically deformable to conventional conductors along
high-temperature paths and all order magnetically at zero temperature,
as Baskaran and Anderson concede.  The ordering temperature is
admittedly low, and as-yet undetected in the case of the sulfate,
and {\it why} it is low is indeed the important matter. Low-temperature
magnetic ordering in real materials is easily disrupted by disorder,
of course, so no report of intrinsically disordered spin ground states
can be believed until spin glass behavior is meticulously searched for
and not found.

Baskaran and Anderson are right about the near equivalence of $^3$He
and the cuprates.  Indeed their assertion that ``the Mott
insulator is a form of quantum solid, and the melting transition in
$^3$He is our best example of a Mott transition'' is quite consistent
with my views in being a first-order transition between two ordered
states - the antiferromagnetic crystal and the p-wave superfluid -
potentially analogous to a first-order transition between the
antiferromagnet and the d-wave superconductor in the cuprates.  However
their criticism that ``no critical point ... connects solid and
liquid'' and that this is ``well-known'' is incorrect. The only thing
required is an extra symmetry, such as that proposed by Zhang \cite{z},
that makes the two kinds of order equivalent, for it is the
incompatability of the broken symmetries that requires the transition
to be first-order. It is true that an extra symmetry of this kind
would be physically unnatural, but nothing prevents a physical
first-order transition from being regulated by an ``unphysical''
critical point nearby in Hamiltonian space.  It should also be noted
that the phase diagram of $^3$He is more complicated than that of the
cuprates, so there are more competing phases to reconcile.

The most important matter raised by Baskaran and Anderson is the
meaning of the fermi surface. The quasiparticle spectrum of the
cuprates in extreme underdoping, i.e. near the ostensible critical
point, develops a ``pseudogap'' of order J, that is 10 times T$_c$,
near the Brillouin zone face at $(\pi , 0)$, and evolves smoothly
into the insulating state, where is possess a deep, isotropic minimum
at $(\pi/2 , \pi/2)$ predicted by a number of us on the basis of
``relativistic'' theories lacking a fermi surface\cite{rbl2}.  So it
is simply not true that ``the low-energy excitations must be described
in terms of a fermi surface''.  I predicted a non-trival photoemission
result quite nicely without it. The fact that the quasiparticle
spectrum evolves continuously from the relativistic behavior at low
doping to the metallic behavior at high doping has the following
simple interpretation:  There are two sets of excitations - one
appropriate to the critical point and one appropriate to the cold
metal - that can used perturbatively to compute measured spectra
across the region of interest.  In either case the ``elementary''
excitations of the perturbation theory scatter strongly at the
temperatures and dopings of interest and lose their integrity as a
result. Neither the fermi surface nor the relativistic point is right;
the whole question of the nature of the elementary excitations is
meaningless because the temperature cannot be lowered to zero.
However, regardless of whether this interpretation is correct it is
experimentally the case the fermi surface loses definition by degrees
as the doping is reduced, and it is therefore not characteristic of
anything.

\begin{appendix}

\section*{Disordered Spin Vacua}

Both the quantum disordered spin vacua discussed in this paper have
prototypes written as projected BCS states at half-filling.  Both may
be generated using fictitious Hamiltonians of the form

\begin{equation}
{\cal H} = \sum_{<jk>} \Psi_j^\dagger \biggl\{ t_{jk} \tau_3
+ \Delta_{jk}^R \tau_1 + \Delta_{jk}^I \tau_2 \biggr\} \Psi_k
\; \; \; ,
\end{equation}

\noindent
where

\begin{displaymath}
\Psi_j = \left[ \begin{array}{c} c_{j \uparrow} \\
         c_{j \downarrow}^\dagger \end{array} \right]
\end{displaymath}

\begin{equation}
\tau_1 = \left[ \begin{array}{rr}
         0 & 1 \\
         1 & 0  \end{array} \right]
\; \; \; \;
\tau_2 = \left[ \begin{array}{rr}
         0 & \! -i \\
         i & 0  \end{array} \right]
\; \; \; \;
\tau_3 = \left[ \begin{array}{rr}
         1 & 0 \\
         0 & \! -1  \end{array} \right]
\; \; \; .
\end{equation}

\noindent
For the specific case of

\begin{displaymath}
t_{jk} = \left[ \begin{array}{rl}
         t & {\rm j \; and \; k \; near \; neighbors} \\
         0 & {\rm otherwise} \end{array} \right]
\end{displaymath}

\begin{displaymath}
\Delta_{jk}^R = \left[ \begin{array}{rl}
           \Delta & {\rm j \; and \; k \; x \; near \; neighbors} \\
          -\Delta & {\rm j \; and \; k \; y \; near \; neighbors} \\
                     0 & {\rm otherwise} \end{array} \right]
\end{displaymath}

\begin{equation}
\Delta_{jk}^I = \left[ \begin{array}{rl}
     \Delta ' & {\rm j \; and \; k \; + \! + \; second \; neighbors} \\
    -\Delta & {\rm j \; and \; k \; + \! - \; second \; neighbors} \\
          0 & {\rm otherwise} \end{array} \right]
\end{equation}

\noindent
we have

\begin{equation}
\Psi_q = \sum_j \exp (i \vec{q} \cdot \vec{r}_j ) \; \Psi_j
\end{equation}

\begin{equation}
{\cal H} = \sum_q \Psi_q^\dagger {\cal H}_q \Psi_q
\end{equation}

\begin{displaymath}
{\cal H}_q = 2 t \biggl[ \cos(q_x) + \cos(q_y) \biggr] \tau_3
\end{displaymath}

\begin{equation}
+ 2 \Delta \biggl[ \cos(q_x) - \cos(q_y) \biggr] \tau_1
+ 4 \Delta ' \sin(q_x) \sin(q_y) \tau_2
\end{equation}

\begin{equation}
{\cal H}_q^2 = E_q^2
\end{equation}

\begin{equation}
| \Phi \! > = \prod_q \biggl[ \Psi_q^\dagger \biggl(
\frac{E_q - {\cal H}_q}{2 E_q} \biggr) \Psi_q \biggr] | 0 \! >
\; \; \; .
\end{equation}

\noindent
The ground state $|\Phi \! >$ thus generated represents a
$d_{x^2 - y^2} + i \epsilon d_{xy}$ superconductor with $\epsilon
= 2 \Delta ' / \Delta$.  When acted upon by $P_\alpha$ with $\alpha
= 1$ at half-filling per Eq. (\ref{gutz}) it results in the chiral
spin liquid vacuum $|\Psi \! >$.  The RVB vacuum is the $\epsilon
\rightarrow 0$ limit of this \cite{affleck}.

There is a local SU(2) gauge symmetry contained in this construction
procedure at half-filling and $\alpha = 1$ caused by overcompleteness
of the representation.  If $| \Psi \! > = P_\alpha |\Phi \! >$ for
this case then it is also true that

\begin{equation}
| \Psi \! > = P_\alpha \; U \; | \Phi \! >
\end{equation}

\noindent
where

\begin{equation}
U = \exp\biggl\{ i \sum_j \Psi_j^\dagger (\vec{\theta}_j \cdot
\vec{\tau}) \Psi_j \biggr\}
\end{equation}

\noindent
for any choice of the variables $\vec{\theta}_j$.  Thus specializing
for simplicity to the case of $\Delta = t$ and taking

\begin{equation}
\vec{\theta_j} \cdot \vec{\tau} =
\frac{\pi}{4} \biggl[ 1 - 2 \cos (\pi x_j )
+ \cos(\pi x_j) \cos(\pi y_j) \biggr] \tau_2
\end{equation}

\noindent
we find that

\begin{equation}
U {\cal H} U^\dagger
= \sum_{<jk>} \Psi_j^\dagger \tilde{t}_{jk} \Psi_k \; \; \; ,
\end{equation}

\noindent
where

\begin{displaymath}
\tilde{t}_{jk} = \tau_3 \exp \biggl\{ i \tau_3 \int_j^k \vec{A} \cdot
d\vec{s} \biggr\}
\end{displaymath}

\begin{equation}
\times
\left[ \begin{array}{rl}
           t & {\rm j \; and \; k \; near \; neighbors} \\
  |\Delta '| & {\rm j \; and \; k \; second \; neighbors}
       \end{array} \right]
\end{equation}

\noindent
with $\vec{A} = \pi y \hat{x}$.  This is the Hamiltonian of electrons
moving in a magnetic field of flux $\pi$ per plaquette, i.e. a
quantum hall problem.

The two kinds of ordering along the $\delta = 0$ line we have discussed
correspond to the two distinct ways a mass can be added to the Dirac
spectrum without destroying relativistic invariance.  The eigenvalue
spectrum of ${\cal H}$ described above expressed either as a
$d_{x^2 - y^2} + i \epsilon d_{xy}$ superconductor or as a lattice
Landau level is

\begin{equation}
E_q = \pm 4t \sqrt{\cos^2(q_x) + \cos^2(q_y) + \epsilon^2
\sin^2(q_x) \sin^2(q_y)} \; \; \; .
\end{equation}

\noindent
The parameter $\epsilon$, which measures the amount of P and T
violation is thus one kind of mass.  The other kind corresponds to
a staggered potential.  Thus taking $\epsilon = 0$ we find that the
eigenvalues of

\begin{equation}
{\cal H} = \sum_{<jk>} \Psi_j^\dagger \tilde{t}_{jk} \Psi_k
+ \sum_j \Psi_j^\dagger V_j \Psi_j \; \; \; ,
\end{equation}

\noindent
where $V_j = \pm V_0$, dependeing on whether j is even or odd, are

\begin{equation}
E_q = \pm 4t \sqrt{\cos^2(q_x) + \cos^2(q_y) + m^2}
\end{equation}

\noindent
with $m = (V_0/4t)$.

\begin{figure}
\epsfbox{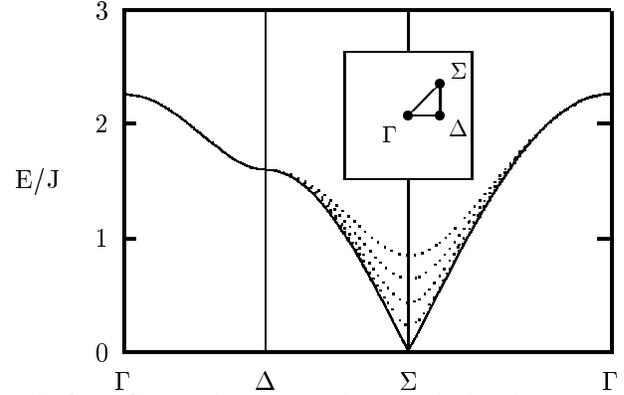}
\caption{Spinon dispersion relation calculated variationally using
         the t-J Hamiltonian and projected particle and hole excitations
         of the $d_{x^2 - y^2}$ superconductor at half-filling.  The
         dashed line shows the opening of the mass gap associated
         with chiral or antiferromagnetic ordering.}
\end{figure}

The particle and hole excitations of the Dirac sea do not correspond
to low-lying excited states under projection except when no
order is developed.  If, for example, the d-wave or flux vacuum
and its particle and hole excitations are used as a variational
ansatz for the t-J model

\begin{equation}
{\cal H}_{tJ} = \sum_{<jk>} \biggl\{ \frac{J}{4} \sum_{\sigma \sigma'}
c_{j\sigma}^\dagger c_{k \sigma'}^\dagger c_{k \sigma} c_{j \sigma'}
+ t \sum_\sigma c_{j\sigma}^\dagger c_{k\sigma} \biggr\}
\end{equation}

\noindent
one obtains

\begin{equation}
E_q^{spinon} = 1.6 J \sqrt{\cos^2(q_x) + \cos^2(q_y)}
\end{equation}

\noindent
for the spinon dispersion relation. This is plotted in Fig. 4. Only
one branch is present because the particle and hole become equivalent
under projection \cite{zou}. Similar considerations applied to the holon
give

\begin{equation}
E_q^{holon} = \pm 2t \sqrt{\cos^2(q_x) + \cos^2(q_y)} \; \; \; .
\end{equation}

\noindent
However, if one uses the correct relaxed vacuum for the t-J model,
which has $\epsilon = 0$ and $m \simeq 0.25$ \cite{hsu}, one finds that
the spinons and holons are no longer free but bind with string
potential \cite{rbl3}. If a Hamiltonian that stabilizes the chiral state
is used, then one finds the potential to be a logarithm.  In any case
attractive interactions between these particles grow with the onset
of order and result in their being bound at low energies except
when the order vanishes.  The functional forms of these potentials
are consistent with the physics of a U(1) gauge theory undergoing a
confinement transition.

\end{appendix}

\end{document}